\documentclass[12pt]{article}
\usepackage{amssymb,amsmath}
\usepackage[dvips]{graphicx,color}
\usepackage{mathrsfs}

\newtheorem{theorem}{Theorem}
\newtheorem{lemma}{Lemma}
\newtheorem{proposition}{Proposition}

\numberwithin{equation}{section}

\date{}
\begin{document}

\title{Local boundary controllability in classes of differentiable functions for the wave equation}
\author{M.I.Belishev\thanks {Saint-Petersburg Department of the Steklov Mathematical Institute, RAS,
                 belishev@pdmi.ras.ru;  Saint-Petersburg State University, 7/9 Universitetskaya nab., St. Petersburg, 199034,
                 Russia, m.belishev@spbu.ru. Supported by the RFBR grant 17-01-00529-à and Volks-Wagen
                 Foundation.}}
\maketitle

\begin{abstract}
The well-known fact following from the Holmgren-John-Tataru
uni\-qu\-e\-ness theorem is a local approximate boundary
$L_2$-controllability of the dynamical system governed by the wave
equation. Generalizing this result, we establish the
controllability in certain classes of differentiable functions in
the domains filled up with waves.
\end{abstract}

\section{Introduction}
The paper deals with a {\it local approximate boundary
controllability} of dy\-na\-mi\-cal systems governed by the wave
equation. This property means that the states of the system
(waves) initiated by the boundary sources (controls), constitute
$L_2$-complete sets in the domains, which the waves fill up. Such
a result is derived from the fundamental Holmgren-John-Tataru
uni\-qu\-e\-ness theorem \cite{Tat} by the scheme proposed by
D.L.Russel \cite{Russell}. The $L_2$-controllability is a
cornerstone of the boundary control method (BC-method), which is
an approach to inverse problems based upon their relations to
control and system theory \cite{BIP97, BIP07}.

In this paper, we show that completeness of waves also holds in
the certain classes of differentiable functions.

The first version of the paper was posted as a preprint
\cite{BDolg} based on the graduate diploma project of
A.N.Dolgoborodov, which was fulfilled under tutorship of the
author in 1997 at the physical faculty of the St-Petersburg State
University. However, it was never published in official issues.
The given variant is a revised and extended version of
\cite{BDolg}. Recently, prof. G.Nakamura informed the author about
certain interest to this kind of results. It is the reason, which
has stimulated to return to this subject.

I'm grateful to A.I.Nazarov for helpful consultations.

\section{Dynamical systems}
All the function classes and spaces are real.
\subsubsection*{Initial boundary value problem}
Let $\Omega\subset{{\mathbb R}^n}$ be a bounded domain with the
$C^\infty$-smooth boundary $\Gamma$;
$T>0,\,\,Q^T:=\Omega\times(0,T),\,\,{\Sigma}^T:=\Gamma\times[0,T]$.
Consider the problem
 \begin{align}
\label{P1} & u_{tt}+Au=0 &&{\rm in}\,\,\,\,Q^T\\
\label{P2} & u|_{t=0}=u_t|_{t=0} &&{\rm in}\,\,\,\,\overline\Omega\\
\label{P3} & u|_{\Sigma^T}=f\,,
 \end{align}
where  $f$ is a {\it boundary control}, $A$ is a differential
expression of the form
 $$
A\,=\,-\sum\limits_{i,j=1}^n\partial_{x^i}a^{ij}(x)\partial_{x^j}
 $$
with the coefficients $a^{ij} \in C^\infty(\overline\Omega)$
provided
 \begin{equation}\label{Eq ellipticity conditions}
a^{ij}(x)=a^{ji}(x)\,;\,\,\, \sum\limits_{i,j=1}^n
a^{ij}(x)\xi_i\xi_j\geqslant \mu |\xi|^2,\quad
x\in\overline\Omega,\,\,\xi=\{\xi_1,\dots, \xi_n\}\in{{\mathbb
R}^n}
 \end{equation}
with a constant $\mu>0$. Let $u=u^f(x,t)$ be a solution ({\it
wave}); the following is the list of its known properties.
\smallskip

(i)\,\,\,Let
 $$
{{\mathscr M}^T}\,:=\,\left\{f\in C^\infty(\Sigma^T)\,|\,\,{\rm
supp\,} f\subset\Gamma\times(0,T]\right\}
 $$
be the class of smooth controls vanishing near $t=0$. This class
is dense in $L_2(\Sigma^T)$. For $f\in{{\mathscr M}^T}$, problem
(\ref{P1})--(\ref{P3}) has a unique classical solution $u^f\in
C^\infty(\overline{Q^T})$. Since the operator $A$, which governs
the evolution of waves, doesn't depend on $t$ and
$\partial_t{\mathscr M}^T={\mathscr M}^T$, this solution satisfies
 \begin{equation}\label{Eq u^f{tt}=-Uu^f}
u^{f_{t}}\,=\,u^f_{t}\,; \qquad
u^{f_{tt}}\,=\,u^f_{tt}\,\overset{(\ref{P1})}=\,-Au^f\,.
 \end{equation}

(ii)\,\,\,The map $f\mapsto u^f$ defined on ${{\mathscr M}^T}$ is
continuous from $L_2(\Sigma^T)$ to $C([0,T]; L_2(\Omega))$\,\,
\cite{LLT}. As such, it can be extended onto $L_2(\Sigma^T)$. From
this point on, for $f\in L_2(\Sigma^T)$, we define a (generalized)
solution $u^f$ of the class $C([0,T]; L_2(\Omega))$ as the image
of $f$ via the extended map.
\smallskip

(iii)\,\,\,By $H^s(...)$ we denote the Sobolev classes. For $f\in
H^{2p}(\Sigma^T)$ provided $(\partial_t)^jf|_{t=0},\,\,j=0,1,\dots
, 2p-1$, one has $u^f\in C([0,T]; H^{2p}(\Omega))$\,\,\cite{LLT,
LTrDynRep}.
\smallskip

(iv)\,\,\,The inverse matrix $\{a_{ij}\}:=\{a^{ij}\}^{-1}$
determines a Riemannian metric $d\tau^2=\sum_{ij}a_{ij}dx^i dx^j$
in $\Omega$ and the corresponding distance ${\rm dist}_A$. Denote
$\tau(x):={\rm dist}_A(x,\Gamma)$ and
 $$
\Omega^r:=\{x\in\overline\Omega\,|\,\,\tau(x)<r\},\,\,\,\,r>0\,.
 $$
The well-known finiteness of the domain of influence principle for
the hyperbolic problem (\ref{P1})--(\ref{P3}) holds and implies
the following equivalent relations
 \begin{equation}\label{Eq supp u^f}
{\rm supp\,} u^f(\cdot,t)\subset
\overline{\Omega^t},\,\,\,\,t>0\,;\quad {\rm supp\,} u^f\subset
\{(x,t)\in \overline{Q^T}\,|\,\,t\geqslant \tau(x)\}
 \end{equation}
(see, e.g., \cite{Ikawa}). So, $\Omega^T$ is the subdomain filled
with waves at the final moment $t=T$. Under our assumptions on
$\Omega$ and ${a^{ij}}$, the value
 $$
T_{\rm fill}\,:=\,\inf\,\{T>0\,|\,\,\Omega^T=\Omega\}
 $$
is finite; we call it a {\it filling time}.

\subsubsection*{Dual problem}
The problem
\begin{align}
\label{D1} & v_{tt}+Av=0 &&{\rm in}\,\,\,\,Q^T\\
\label{D2} & v|_{t=T}=0,\,\,\,v_t|_{t=T}=y &&{\rm in}\,\,\,\,\overline\Omega\\
\label{D3} & v|_{\Sigma^T}=0
 \end{align}
is called {\it dual} to problem (\ref{P1})--(\ref{P3}); let
$v=v^y(x,t)$ be its solution. The following is the list of its
known properties.
\smallskip

(i*)\,\,\,For $y\in C^\infty_0(\Omega)$, problem
(\ref{D1})--(\ref{D3}) has a unique classical solution $v^y\in
C^\infty(\overline{Q^T})$. The map $y\mapsto v^y$ acts
continuously from $L_2(\Omega)$ to $C([0,T];
H^1_0(\Omega))$\,\,\cite{LLT, LTrDynRep}.
\smallskip

(ii*)\,\,\,Let $\partial_{\nu_A}:=\sum_{i,j=1}^n a^{ij}\cos(\nu,
x^j)\partial_{x^i}$ be the conormal derivative at the boundary
$\Gamma$ \,(here $\nu$ is the Euclidean normal). The map $y\mapsto
\partial_{\nu_A}v^y|_{\Sigma^T}$ is continuous from $L_2(\Omega)$ to
$L_2(\Sigma^T)$\,\,\cite{LLT, LTrDynRep}.
\smallskip

(iii*)\,\,\,By the finiteness of the domain of influence principle
for the hyperbolic problem (\ref{D1})--(\ref{D3}), the trace
$\partial_{\nu_A}v^y|_{\Sigma^T}$ is determined by the values of
$y|_{\Omega^T}$ (does not depend on
$y|_{\Omega\setminus\Omega^T}$). In particular, if
$y|_{\Omega^T}=0$ then $\partial_{\nu_A}v^y|_{\Sigma^T}=0$ holds.
\smallskip

\subsubsection*{Spaces and operators}
Here we consider the above introduced problems as dynamical
systems and endow them with the standard attributes of control and
system theory.
\smallskip

\noindent$\bullet$\,\,\,The Hilbert space of controls ${\mathscr
F}^T:=L_2(\Sigma^T)$ is an {\it outer} space of the system
(\ref{P1})--(\ref{P3}).

The Hilbert space ${\mathscr H}:=L_2(\Omega)$ is called an {\it
inner} space. It contains the subspace ${{\mathscr
H}^T}:=\{y\in{\mathscr H}\,|\,\,{\rm supp\,}
y\subset\overline{\Omega^T}\}$ of functions supported in the
subdomain filled up with waves at the final moment $t=T$.
\smallskip

\noindent$\bullet$\,\,\,The map $W^T:{\mathscr F}^T\to{\mathscr
H},\,\,\,W^Tf:=u^f(\cdot,T)$ is a {\it control operator}. By the
property (ii), it is continuous.

The map $O^T:{\mathscr H}\to{\mathscr
F}^T,\,\,\,O^Ty:=\partial_{\nu_A}v^y|_{\Sigma^T}$ associated with
the system (\ref{D1})--(\ref{D3}) is an {\it observation
operator}. The well-known fact is the duality relation
 \begin{equation}\label{Eq O=W^*}
O^T\,=\,(W^T)^*,
 \end{equation}
which is derived by integration by parts (see, e.g., \cite{BIP97,
BIP07}).
\smallskip

\noindent$\bullet$\,\,\,The set of waves
 \begin{equation}\label{Eq def U^T}
{\mathscr U}^T:=\{u^f(\cdot,T)\,|\,\,f\in {\mathscr
F}^T\}=W^T{\mathscr F}^T={\rm Ran\,} W^T\subset{\mathscr H}
 \end{equation}
is called {\it reachable} (at the moment $t=T$). By the first
relation in (\ref{Eq supp u^f}), the embedding
 \begin{equation}\label{Eq U^T subset H^T}
{\mathscr U}^T\,\subset {{\mathscr H}^T}\,,\qquad T>0
 \end{equation}
holds. The general operator equality implies
 $$
{\rm Ker\,} O^T={\mathscr H}\ominus\overline{{\rm Ran\,}
(O^T)^*}\overset{(\ref{Eq O=W^*})}={\mathscr
H}\ominus\overline{{\rm Ran\,} W^T}={\mathscr
H}\ominus\overline{{\mathscr U}^T}
 $$
(see, e.g., \cite{BirSol}), whereas (\ref{Eq U^T subset H^T})
leads to
 \begin{equation}\label{Eq Ker O subset H-H^T subset H^T}
{\rm Ker\,} O^T\,\supset\,{\mathscr H}\ominus{{\mathscr
H}^T}\,,\qquad T>0
 \end{equation}
that corresponds to the property (iii*).

\section{Controlability}
\subsubsection*{$L_2$-controllability}
One of the central results of the boundary control theory, which
plays the crucial role for the BC-method, is that the embedding
(\ref{Eq U^T subset H^T}) is dense:
 \begin{equation}\label{Eq U^T = H^T}
\overline{{\mathscr U}^T}\,=\,{{\mathscr H}^T}\,,\qquad T>0
 \end{equation}
(see \cite{BIP97, BIP07}). In particular, for $T>T_{\rm fill}$ one
has $\overline{{\mathscr U}^T}\,=\,{\mathscr H}$. As was mentioned
in Introduction, (\ref{Eq U^T = H^T}) is derived from the
Holmgren-John-Tataru Theorem on uniqueness of continuation of the
solutions to the wave equation across a noncharacteristic surfaces
\cite{Tat}. This result means that any function supported in the
domain $\Omega^T$ filled with waves can be approximated (in the
$L_2$-metric) by a wave $u^f(\cdot,T)$ with the properly chosen
control $f\in{\mathscr F}^T$. In control theory such a property is
referred to as a {\it local approximate boundary controllability}
of system (\ref{P1})--(\ref{P3}).
\smallskip

Since ${\rm Ker\,} O^T={\mathscr H}\ominus\overline{{\mathscr
U}^T}$, property (\ref{Eq U^T = H^T}) leads to the equality
 \begin{equation}\label{Eq Ker O^T = H-H^T}
{\rm Ker\,} O^T\,=\,{\mathscr H}\ominus{{\mathscr H}^T}\,,\qquad
T>0\,,
 \end{equation}
which refines (\ref{Eq Ker O subset H-H^T subset H^T}) and is
interpreted as an {\it observability} of the dual system
(\ref{D1})--(\ref{D3}). It means that the wave $v^y$ isn't
observed at the boundary during the interval $0\leqslant
t\leqslant T$ if and {\it only if} the velocity perturbation $y$,
which initiates the wave process, is separated from the boundary:
${\rm dist}_A({\rm supp\,} y,\Gamma)\geqslant T$. In particular,
for $T>T_{\rm fill}$ one has ${\rm Ker\,}O^T=\{0\}$. The duality
`controllability--observability' is a very general fact of the
system theory.
\smallskip

Later on we'll use the following quite evident consequence of the
observability (\ref{Eq Ker O^T = H-H^T}).
 \begin{proposition}\label{Prop1}
Let $0<\delta<T$. The relation $O^Ty|_{\Gamma\times[\delta,T]}=0$
implies $y\in{\mathscr H}\ominus{\mathscr H}^{T-\delta}$ that is
equivalent to ${\rm supp\,}y\subset
\Omega\setminus\Omega^{T-\delta}$.
 \end{proposition}

\subsubsection*{Spaces ${\mathscr D}_s$}
As is well known, the operator
 $$
A_0: {\mathscr H}\to{\mathscr H},\,\,\,{\rm Dom\,}
A_0=H^2(\Omega)\cap H^1_0(\Omega),\,\,\, A_0 y\,:=\,A y
 $$
is positive definite in ${\mathscr H}$ and has a purely discrete
spectrum $\{\lambda_k\}_{k\geqslant 1}:
\,0<\lambda_1<\lambda_2\leqslant\lambda_3\leqslant\dots;\,\,\lambda_k\to
\infty$. Let $\{e_k\}_{k\geqslant 1}: A_0 e_k=\lambda_ke_k$ be the
basis of its eigenfunctions normalized by $(e_k,e_l)_{\mathscr
H}=\delta_{kl}$.
\smallskip

For $s>0$, define the Hilbert space of functions
 $$
{\mathscr D}_s\,:=\,{\rm Dom\,}A^{s/2}_0\,,\qquad (y,w)_{{\mathscr
D}_s}\,:=\,(A^{s/2}_0 y, A^{s/2}_0 w)_{{\mathscr H}}
 $$
and note the relations ${\mathscr D}_s\subset H^s(\Omega)$ and
${\mathscr D}_s\supset{\mathscr D}_{s'}$ as $s<s'$ (see
\cite{LM}). This space contains the subspace
 \begin{equation}\label{Eq def D^Ts}
{\mathscr D}_s^T\,:=\,\overline{\{y\in{\mathscr D}_s\,|\,\,{\rm
supp\,}y\subset{\Omega^T}\cup\Gamma\,\}}
 \end{equation}
of functions supported in the filled domain. The definition easily
implies
\begin{equation}\label{Eq DT=cup D{T-delta}}
{\mathscr D}_s^T\,=\,\overline{\bigcup_{0<\delta<T}{\mathscr
D}_s^{T-\delta}}\,.
 \end{equation}
\smallskip

Introduce the (sub)class of smooth controls
 $$
{\mathscr M}^T_0\,:=\,\{f\in {\mathscr
M}\,|\,\,\,\,\partial^{2p}_tf|_{t=T}=0,\,\,p=0,1,2,\dots\}
 $$
and note that $\partial^{2p}_t{\mathscr M}^T_0\subset{\mathscr
M}^T_0$ holds for all $p\geqslant 0$. Let
 $$
{\mathscr U}^T_0\,:=\,\{u^f(\cdot,T)\,|\,\,\,f \in {\mathscr
M}^T_0\}\,=\,W^T{\mathscr M}^T_0
 $$
be the corresponding reachable set.
 \begin{proposition}\label{Prop2}
The embedding ${\mathscr U}^T_0\subset{\mathscr D}^T_s$ holds for
all $s>0$ and $T>0$..
 \end{proposition}
Indeed, if $f\in{\mathscr M}^T_0$ then $u^f(\cdot,T)\in
C^\infty(\overline\Omega)$,
$u^f(\cdot,T)|_\Gamma\overset{(\ref{P3})}=f|_{t=T}=0$ and, hence,
$u^f(\cdot,T)\in{\rm Dom\,}A_0$. Therefore,
 $$
A_0u^f(\cdot,T)=Au^f(\cdot,T)\overset{(\ref{Eq
u^f{tt}=-Uu^f})}=-u^{f_{tt}}(\cdot,T)\in{\rm Dom\,}A_0
 $$
since $f_{tt}\in{\mathscr M}^T_0$. Thus, we have
$u^f(\cdot,T)\in{\rm Dom\,}A_0^2$. Going on in the evident way, we
get $u^f(\cdot,T)\in{\rm Dom\,}A_0^p$ with any integer $p\geqslant
1$. Hence, $u^f(\cdot,T)\in{\mathscr D}_s$ for all $s>0$. In the
mean time, ${\rm supp\,}u^f(\cdot,T)\subset\Omega^T\cup\Gamma$
and, hence, $u^f(\cdot,T)\in{\mathscr D}_s^T$. The Proposition is
proven.

\subsubsection*{${\mathscr D}_s$-controllability}
The following result is referred to as an approximate boundary
${\mathscr D}_s$-controll\-abi\-li\-ty of system
(\ref{P1})--(\ref{P3}).
 \begin{theorem}\label{Th1}
The relation
\begin{equation}\label{Eq U0^T = Ds^T}
\overline{{\mathscr U}^T_0}\,=\,{\mathscr D}^T_s\,,\qquad
s>0,\,\,\,T>0
 \end{equation}
(the closure in ${\mathscr D}_s$) is valid.  In particular, for
$T>T_{\rm fill}$ one has $\overline{{\mathscr
U}^T_0}\,=\,{\mathscr D}_s$.
 \end{theorem}
{\bf Proof.}

{\bf 1.}\,\,\,{\it Spectral representation.}\,\,\,Recall that
$\{\lambda_k\}_{k\geqslant 1}$ and $\{e_k\}_{k\geqslant 1}$ are
the spectrum and basis (in $\mathscr H$) of eigenfunctions of the
operator $A_0$. As is easy to check, the system
 $$
\{e^s_k\}_{k\geqslant 1}:\,\,\,\,\,e^s_k:= \lambda_k^{-{s/2}}e_k
 $$
constitutes an orthogonal normalized basis in ${\mathscr D}_s$.

The system
 \begin{align}
\label{DD1} & v_{tt}+Av=0 &&{\rm in}\,\,\,\,\Omega\times{\mathbb R}\\
\label{DD2} & v|_{t=T}=0,\,\,\,v_t|_{t=T}=y &&{\rm in}\,\,\,\,\overline\Omega\\
\label{DD3} & v|_{\Gamma\times{\mathbb R}}=0
 \end{align}
is an extension of the dual system to all times. For a $y\in
C^\infty_0(\Omega)$ it has a unique classical solution $v^y\in
C^\infty({\overline\Omega}\times{\mathbb R})$. Applying the
Fourier method to problem (\ref{D1})--(\ref{D3}), one easily
derives
 \begin{equation}\label{Eq v^y spectral expansion}
v^y(\cdot,t)\,=\,\sum\limits_{k=1}^\infty
\alpha_k\,\frac{\sin\sqrt{\lambda_k}(t-T)}{\sqrt{\lambda_k}}\,e_k\,;\qquad
\alpha_k=(y,e_k)_{\mathscr H}\,.
 \end{equation}
Note that $v^y(\cdot,t)$ is odd w.r.t. $t=T$.

{\bf 2.}\,\,\,{\it Regularization.}\,\,\,For an arbitrary $y
\in{\mathscr H}$, the (generalized) solution $v^y(\cdot,t)$ is
also represented by the right hand side of (\ref{Eq v^y spectral
expansion}) but may not belong to the classes ${\mathscr D}_s$.
Here we provide a procedure, which improves smo\-oth\-ness of
solutions to the dual system.

\noindent$\bullet$\,\,\,The role of the smoothing kernel is played
by a function
 $$
\phi_\varepsilon(t)\,:=\,\varepsilon^{-1}\phi(\varepsilon^{-1}t)\,\underset{\varepsilon
\to +0}\to\,\delta(t)\,,
 $$
where $\phi\in C^\infty({\mathbb R})$ satisfies
 $$
\phi\geqslant 0; \quad \phi(-t)=\phi(t);\quad {\rm
supp\,}\phi\subset[-1,1];\quad\int_{\mathbb R}\phi(t)\,dt\,=1\,.
 $$

Let $y\in C^\infty_0(\Omega)$. The function
 \begin{equation}\label{Eq v^y eps def}
v^y_\varepsilon(\cdot,t)\,:=\,[\phi_\varepsilon \ast
v^y](\cdot,t)\,=\,\int_{\mathbb
R}\phi_\varepsilon(\eta)v^y(\cdot,t-\eta)\,d\eta\,,\qquad
t\in\mathbb R
 \end{equation}
(the convolution w.r.t. time) is also $C^\infty$-smooth in
${\overline\Omega}\times{\mathbb R}$ and odd w.r.t. $t=T$.
Integrating in (\ref{Eq v^y spectral expansion}), one easily gets
its spectral representation:
\begin{align}
\notag & v^y_\varepsilon(\cdot,t)\,=\,\sum\limits_{k=1}^\infty
\beta_k^\varepsilon
\alpha_k\,\frac{\sin\sqrt{\lambda_k}(t-T)}{\sqrt{\lambda_k}}\,e_k\,;\qquad
\alpha_k=(y,e_k)_{\mathscr H}\,,\\
\label{Eq v^y eps spectral expansion} &
\beta_k^\varepsilon=\int_{-\varepsilon}^\varepsilon\phi_\varepsilon(\eta)\,\cos\sqrt{\lambda_k}\eta\,d\eta=
\int_{-1}^1\phi(t)\,\cos\varepsilon\sqrt{\lambda_k}t\,dt
 \end{align}
with $|\beta_k^\varepsilon|\leqslant 1$. Taking into account the
properties of $\phi$, one can easily derive
 \begin{equation}\label{Eq alpa beta assympt}
\beta_k^\varepsilon\underset{\varepsilon\to 0}\to
1\,,\,\,\,\,\,\,k\geqslant 1;\qquad \lambda^{s/2}_k
\beta_k^\varepsilon\underset{k\to \infty}\to
0\,,\,\,\,\,\,s\geqslant0,\,\,\varepsilon>0\,.
 \end{equation}

Comparing (\ref{Eq v^y spectral expansion}) with (\ref{Eq v^y eps
spectral expansion}), we see that $v^y_\varepsilon$ is a solution
to problem (\ref{DD1})--(\ref{DD3}) satisfying
 \begin{equation}\label{Eq v^y eps for t=T}
v^y_\varepsilon|_{t=T}=0\,, \quad
(v^y_\varepsilon)_t|_{t=T}\,=\,\sum\limits_{k=1}^\infty
\beta_k^\varepsilon \alpha_k\,e_k\,=:\,y_\varepsilon\,.
 \end{equation}
So, we have $v^y_\varepsilon=v^{y_\varepsilon}$. Also, note that
$y_\varepsilon \in{\mathscr D}^s$ for all $s>0$ by virtue of the
second relation in (\ref{Eq alpa beta assympt}).
\smallskip

\noindent$\bullet$\,\,\,By the aforesaid, the operator ({\it
regularizer}) $R_\varepsilon: {\mathscr H}\to{\mathscr
H},\,\,R_\varepsilon y:=y_\varepsilon$ is well defined on
$C^\infty_0(\Omega)$. Estimating
 $$
\|y_\varepsilon\|^2_{{\mathscr D}_s}\,=\,\sum\limits_{k=1}^\infty
\lambda_k^s (\beta_k^\varepsilon)^2 \alpha_k^2\overset{(\ref{Eq
alpa beta assympt})}\leqslant\,{\rm
const\,}\sum\limits_{k=1}^\infty \alpha_k^2\,={\rm
const\,}\|y\|_{\mathscr H}^2\qquad(\varepsilon>0)\,,
 $$
we see that $R_\varepsilon$ acts continuously from ${\mathscr H}$
to ${\mathscr D}_s$.

Representation (\ref{Eq v^y eps for t=T}) implies
 \begin{equation}\label{Eq Re=beta e}
R_\varepsilon e_k\,=\,\beta_k^\varepsilon\,e_k\,,
 \end{equation}
i.e., the regularizer is diagonal in the eigenbasis of $A_0$.
Hence, we have $R_\varepsilon
e_k^s\,=\,\beta_k^\varepsilon\,e_k^s$ in ${\mathscr D}_s$. Since
$\beta_k^\varepsilon$ are uniformly bounded, $R_\varepsilon$ is
continuous as an operator in ${\mathscr D}_s$ and its norm is
bounded uniformly w.r.t. $\varepsilon$. In the mean time, by the
first relation in (\ref{Eq alpa beta assympt}), the regularizer
converges to the identical operator $I$ on the dense set ${\rm
span\,}\{e_k^s\}_{k\geqslant 1}$ as $\varepsilon\to 0$. As a
result, the convergence $R_\varepsilon\underset{\varepsilon\to
0}\to I$ in the strong operator topology in ${\mathscr D}_s$ does
occur.

\noindent$\bullet$\,\,\,Fix $\delta\in (0,T)$ and a positive
$\varepsilon<\delta$. For the controls $f\in{\mathscr F}^T$
provided ${\rm supp\,}f\subset \Gamma\times[\delta,T]$, the
operation $f\,\mapsto\,f_\varepsilon$:
 \begin{equation}\label{Eq def f eps}
f_\varepsilon(\cdot,t):=\int_0^T[\phi_\varepsilon(t-\eta)-\phi_\varepsilon(2T-t-\eta)]\,f(\cdot,\eta)\,d\eta,
\quad 0\leqslant t\leqslant T
 \end{equation}
is well defined. With regard to the definition of the class
${\mathscr M}^T_0$ and properties of the kernel
$\phi_\varepsilon$, one can easily check that
$f_\varepsilon\in{\mathscr M}^T_0$. Note that the latter implies
 $$
W^Tf_\varepsilon=u^{f_\varepsilon}(\cdot,T)\in {\mathscr
U}^T_0\subset{\mathscr D}^T_s\,.
 $$
 \begin{proposition}\label{Prop3}
For any admissible $f\in{\mathscr F}^T$ and $y\in{\mathscr H}$,
the relation
 \begin{equation}\label{Eq O^T and R eps}
(f_\varepsilon, O^T y)_{{\mathscr F}^T}\,=\,(f, O^T R_\varepsilon
y)_{{\mathscr F}^T}
 \end{equation}
holds.
 \end{proposition}
Indeed, let $y\in C^\infty_0(\Omega)$, so that $v^y$ is classical
and smooth in $\overline\Omega\times \mathbb R$.  Applying
$\partial_{\nu_A}$ in (\ref{Eq v^y eps def}), we get
 $$
\partial_{\nu_A}v^y_\varepsilon\,=\,\phi_\varepsilon \ast
\partial_{\nu_A}v^y\qquad {\rm on}\,\,\,\Gamma\times{\mathbb R}\,.
 $$
By the evenness/oddness of $\phi_\varepsilon$ and $v^y$, for the
times $t<T$ the right hand side can be written in the form
 \begin{equation}\label{Eq auxill}
[\phi_\varepsilon \ast
\partial_{\nu_A}v^y](\cdot,t)\,=\,\int_{-\infty}^T[\phi_\varepsilon(t-\eta)-\phi_\varepsilon(2T-t-\eta)]\,
\partial_{\nu_A}v^y(\cdot,\eta)\,d\eta\,.
 \end{equation}
Taking into account (\ref{Eq def f eps}), (\ref{Eq auxill}) and
changing the order of integration, one derives
 \begin{align*}
& (f_\varepsilon, O^T y)_{{\mathscr
F}^T}\,=\,\int_{\Sigma^T}f_\varepsilon(\gamma,t)\,\partial_{\nu_A}v^y(\gamma,t)\,d\Gamma
dt=\\
&
=\,\int_{\Sigma^T}f(\gamma,t)\left[\int_{-\varepsilon}^\varepsilon\phi_\varepsilon(\eta)
\partial_{\nu_A}v^y(\gamma,t-\eta)\,d\eta\right]d\Gamma dt=\\
& =
\int_{\Sigma^T}f(\gamma,t)\partial_{\nu_A}v^y_\varepsilon(\gamma,t-\eta)\,d\Gamma
dt\,=\\
&
=\,\int_{\Sigma^T}f(\gamma,t)\partial_{\nu_A}v^{y_\varepsilon}(\gamma,t-\eta)\,d\Gamma
dt\,=\,(f, O^T R_\varepsilon y)_{{\mathscr F}^T}\,.
 \end{align*}
Thus, we get (\ref{Eq O^T and R eps}) for the given $y$. Since
such $y$'s constitute a dense set in ${\mathscr H}$, whereas the
operator $O^TR_\varepsilon$ is continuous, we extend (\ref{Eq O^T
and R eps}) to all $y\in{\mathscr H}$. The Proposition is proven.
\smallskip

{\bf 3.}\,\,\,{\it Completing the proof of Theorem.}\,\,\,Let
$z\in{\mathscr D}^T_s\ominus\overline{{\mathscr U}^T_0}$ (the
orthogonality in ${\mathscr D}_s$); we are going to show that
$z=0$. Recall that the scalar product in ${\mathscr D}_s$ is
 $$
(y,w)_{{\mathscr D}_s}\,=(A^{s/2}_0 y, A^{s/2}_0 w)_{\mathscr
H}\,=\,\,\sum\limits_{k=1}^\infty\lambda^s_k\,(y,e_k)_{\mathscr
H}(w,e_k)_{\mathscr H}\,.
 $$

Let $\alpha_k=(z,e_k)_{\mathscr H}$. Fix a $\delta\in(0,T)$ and
positive $\varepsilon<\delta$. By the choice of $z$, for
$f\in{\mathscr F}^T$ provided ${\rm
supp\,}f\subset\Gamma\times[\delta,T]$ one derives
 \begin{align*}
& 0\,=\,(W^Tf_\varepsilon, z)_{{\mathscr
D}_s}\,=\,\sum\limits_{k=1}^\infty\lambda^s_k\,(W^Tf_\varepsilon,e_k)_{\mathscr
H}(z,e_k)_{\mathscr H}\overset{(\ref{Eq
O=W^*})}=\,\sum\limits_{k=1}^\infty\lambda^s_k\,\alpha_k(f_\varepsilon,O^Te_k)_{{\mathscr F}^T}\,=\\
& \overset{(\ref{Eq O^T and R
eps})}=\,\sum\limits_{k=1}^\infty\lambda^s_k\,\alpha_k(f,O^TR_\varepsilon
e_k)_{{\mathscr F}^T}\,\overset{(\ref{Eq Re=beta
e})}=\,\sum\limits_{k=1}^\infty\lambda^s_k\,\alpha_k\beta_k^s\,(f,O^T
e_k)_{{\mathscr F}^T}\,=\\
&=\,\left(f,\,\,
O^T\sum\limits_{k=1}^\infty\lambda^s_k\,\alpha_k\beta_k^s\,
e_k\right)_{{\mathscr F}^T}\,\overset{(\ref{Eq v^y eps for t=T})}=\,(f, O^T A^s_0 z_\varepsilon)_{{\mathscr F}^T}\,,\\
 \end{align*}
the continuity of $O^T:{\mathscr H}\to{\mathscr F}^T$ being in the
use. Since $f$ is arbitrary on $\Gamma\times[\delta,T]$, we see
that
 $$
(O^T A^s_0 z_\varepsilon)\big|_{\Gamma\times[\delta,T]}\,=\,0\,.
 $$
By Proposition \ref{Prop1}, the latter implies
 \begin{equation}\label{Eq A^S0 z eps=0}
A^s_0 z_\varepsilon\,=\,0 \qquad {\rm
in}\,\,\,\Omega^{T-\delta}\,.
 \end{equation}

Next, for a $y\in {\mathscr D}_s^{T-\delta}$ we have
 $$
(y, z_\varepsilon)_{{\mathscr D}_s}\,=\,(A^{s/2}_0y, A^{s/2}_0
z_\varepsilon)_{\mathscr H}\,=\,(y, A^{s}_0
z_\varepsilon)_{\mathscr H}\,\overset{(\ref{Eq A^S0 z
eps=0})}=\,0\,,
 $$
i.e., $z_\varepsilon\in{\mathscr D}^T_s\ominus{\mathscr
D}_s^{T-\delta}$. Tending $\varepsilon\to0$, we get
$z_\varepsilon=R_\varepsilon z\to z$ in ${\mathscr D}_s$ and
conclude that $(y, z)_{{\mathscr D}_s}=0$ for any $y\in {\mathscr
D}_s^{T-\delta}$ and $\delta\in(0,T)$. Referring to (\ref{Eq
DT=cup D{T-delta}}), we arrive at $z=0$ and, thus, prove Theorem
\ref{Th1}.

\subsubsection*{$H^1_0$-controllability}
In the rest of the paper we consider certain applications of
Theorem \ref{Th1}.
\smallskip

In terms of the Riemannian geometry in $\Omega$ determined by the
metric $d\tau^2=a_{ij}\,dx^i dx^j$, the subdomain $\Omega^T$ is a
near-boundary layer of the thickness $T$. It increases as $T$
grows. Recall that $\tau(x)={\rm dist}_A(x,\Gamma)$. For $T<T_{\rm
fill}$, the boundary of the layer consists of two parts:
$\partial\Omega^T=\Gamma\cup\Gamma^T$, where
 $$
\Gamma^T\,:=\,\{x\in\Omega\,|\,\,\tau(x)=T\}
 $$
is a surface equidistant to $\Gamma$.
\smallskip

The smoothness of $\Gamma$ provides ${\mathscr
D}_1=H^1_0(\Omega)$, i.e., these spaces consist of the same
reserve of functions, whereas the norms $\|\cdot\|_{{\mathscr
D}_1}$ and $\|\cdot\|_{H^1_0(\Omega)}$ are equi\-valent. Hence,
one has
 $$
{\mathscr D}^T_1\overset{(\ref{Eq def D^Ts})}=\,\overline{\{y\in
H^1_0(\Omega)\,|\,\,\,{\rm supp\,}y\subset\Omega^T\cup\Gamma)\}}\,=\,H^1_0(\Omega^T)\\
 $$
(the closure in $H^1$-metric), the latter equality being valid
since the compactly supported functions are dense in
$H^1_0(\Omega^T)$. As a result, we arrive at
 \begin{equation}\label{Eq H^10 controllability}
\overline{{\mathscr U}^T_0}\,=\,H^1_0(\Omega^T)\,,\,\,\qquad
T>0\,.
 \end{equation}

\subsubsection*{$H^1$-controllability}
Using the wider class of controls ${\mathscr M}^T$ instead of
${\mathscr M}^T_0$, one extends the corres\-ponding reachable set
from ${\mathscr U}^T_0$ to
 $$
{\mathscr U}^T_*:=\{u^f(\cdot,T)\,|\,\,f \in {\mathscr
M}^T\}\,=\,W^T{\mathscr M}^T\,,
 $$
so that ${\mathscr U}^T_0\subset{\mathscr U}^T_*\subset{\mathscr
U}^T$ holds.

For $T>0$, define the class
 \begin{equation*}
H^1_*(\Omega^T)\,:=\,\overline{\{y\in H^1(\Omega)\,|\,\,{\rm
supp\,}y\subset\Omega^T\cup\Gamma\}}
 \end{equation*}
(the closure in $H^1(\Omega)$). Its elements differ from the ones
of $H^1_0(\Omega)$ by that $y|_\Gamma=0$ is cancelled:
$H^1_0(\Omega)=\{y\in H^1_*(\Omega^T)\,|\,\,\,\,y|_\Gamma=0\}$.
 \begin{lemma}\label{L1}
For any $T>0$, the relation
 \begin{equation*}
\overline{{\mathscr U}^T_*}\,=\,H^1_*(\Omega^T)
 \end{equation*}
(the closure in $H^1(\Omega)$) is valid. In particular, for
$T>T_{\rm fill}$ one has $\overline{{\mathscr
U}^T_*}\,=\,H^1(\Omega)$.
 \end{lemma}
{\bf Proof.}

\noindent$\bullet$\,\,\,The well-known geometric fact, which is
popularly referred to as a variant of the `collar theorem', is
that there exists a domain $\dot\Omega\Supset\Omega$ with the
properties listed below. The objects related with it are marked
with dots.
 \begin{enumerate}
\item The boundary $\partial\dot\Omega=:\dot\Gamma$ is
$C^\infty$-smooth. The coefficients $\dot a^{ij}\in
C^\infty(\,\overline{\dot\Omega}\,)$ obey the ellipticity
conditions (\ref{Eq ellipticity conditions}) with a constant $\dot
\mu>0$ and satisfy $\dot a^{ij}|_{\Omega}= a^{ij}$.

\item The distance ${\rm dist}_{\dot A}$ in $\dot\Omega$ is such
that $\dot\Gamma^\eta=\Gamma$ for some $\eta>0$, and,
respectively, $\dot\Gamma^{T+\eta}=\Gamma^T\,\,\,\,(T>0)$.
 \end{enumerate}
\smallskip

\noindent$\bullet$\,\,\,Let $y\in H^1_*(\Omega^T)$. To prove the
Lemma, it suffices to construct a sequence $\{f_j\}_{j\geqslant
1}\subset{\mathscr M}^T_*$ such that $u^{f_j}(\cdot,T)\to y$ in
$H^1(\Omega)$. We do it as follows.

Extend $y$ to a function $\dot y\in H^1_0(\dot\Omega^{T+\eta}):
\dot y|_{\Omega}=y$. Such an extension does exist owing to
smoothness of $\Gamma$: see, e.g., \cite{LM}.

Consider problem (\ref{P1})--(\ref{P3}) in $\dot\Omega\times
(0,T+\eta)$. By (\ref{Eq H^10 controllability}), one can choose
the controls $\{\dot f_j\}\subset \dot{\mathscr M}^{T+\eta}_0$ so
that $u^{\dot f_j}(\cdot,T+\eta)\to \dot y$ in $H^1(\dot\Omega)$.
Correspondingly, the convergence $u^{\dot
f_j}(\cdot,T+\eta)|_{\Omega}\to \dot y|_\Omega=y$ holds in
$H^1(\Omega)$.
\smallskip

\noindent$\bullet$\,\,\,Return to problem (\ref{P1})--(\ref{P3})
in $\Omega\times (0,T)$ and put $\{f_j\}\subset {\mathscr
F}^T:\,\,\,f_j(\cdot,t):=u^{\dot
f_j}(\cdot,t+\eta)|_\Gamma,\,\,0\leqslant t\leqslant T$. Recalling
the properties (\ref{Eq u^f{tt}=-Uu^f}) and (\ref{Eq supp u^f}),
one can easily verify that $f_j\in{\mathscr M}^T_*$ holds and
provides $u^{f_j}(\cdot,t)=u^{\dot f_j}(\cdot,t+\eta)|_\Omega\to
y$. The Lemma is proven.

\subsubsection*{$H^p$- and $C^m$-controllability}
The result of Lemma \ref{L1} can be easily generalized as follows.
\smallskip

\noindent$\bullet$\,\,\,In the spaces $H^p(\Omega)$ for
$p=0,1,2,\dots$ define the subspaces
 \begin{equation*}
H^p_*(\Omega^T)\,:=\,\overline{\{y\in H^p(\Omega)\,|\,\,{\rm
supp\,}y\subset\Omega^T\cup\Gamma\}}\,, \qquad T>0\,.
 \end{equation*}
The relation
\begin{equation*}
\overline{{\mathscr U}^T_*}\,=\,H^p_*(\Omega^T)\,, \qquad T>0
 \end{equation*}
(the closure in $H^1(\Omega)$) is valid for all $T>0$.
\smallskip

\noindent$\bullet$\,\,\,Let $m=0,1,2,\dots$. In the spaces
$C^m(\overline\Omega)$, define the subspaces
 \begin{equation*}
C^m_*(\Omega^T)\,:=\,\overline{\{y\in
C^m(\overline\Omega)\,|\,\,{\rm
supp\,}y\subset\Omega^T\cup\Gamma\}}\,, \qquad T>0\,.
 \end{equation*}
By the Sobolev embedding theorems, for $s\geqslant
m+1+\left[\frac{n}{2}\right]$ the relation $H^s(\Omega)\subset
C^m(\overline\Omega)$ holds\,\cite{LM}. As a simple consequence,
one has
\begin{equation*}
\overline{{\mathscr U}^T_*}\,=\,C^m_*(\Omega^T)\,, \qquad T>0
 \end{equation*}
(the closure in $C^m(\overline\Omega)$).
\smallskip

\noindent$\bullet$\,\,\, Note in conclusion that all the above
obtained results are valid in the case
$A\,=\,-\sum\limits_{i,j=1}^n\partial_{x^i}a^{ij}(x)\partial_{x^j}+q$
with $q\in C^\infty(\overline\Omega)$.

\end{document}